\documentclass[aps,pra,twocolumn,showpacs,superscriptaddress,longbibliography]{revtex4-2}
\usepackage{graphicx} 
\usepackage{amsmath,amssymb,graphicx}
\usepackage{gensymb}
\usepackage[dvipsnames]{xcolor}

\newcommand{\be}{\begin{equation}}
	\newcommand{\ee}{\end{equation}}
\newcommand{\bea}{\begin{eqnarray}}
	\newcommand{\eea}{\end{eqnarray}}
\newcommand{\bse}{\begin{subequations}}
	\newcommand{\ese}{\end{subequations}}

\usepackage{color}
\usepackage[colorlinks,bookmarks=false,citecolor=darkblue,linkcolor=red,urlcolor=blue]{hyperref}

\definecolor{darkred}{rgb}{0.7,0.0,0.0}

\definecolor{darkblue}{rgb}{0,0.02,0.45}

\definecolor{darkgreen}{rgb}{0.02,0.45,0.0}

\definecolor{violet}{rgb}{0.8,0.2,0.6}

\graphicspath{{figures/}}

\begin{document}
\title{Quantum magnetism of ferromagnetic spin dimers in $\alpha$-KVOPO$_4$}

\author{Prashanta K. Mukharjee}
\thanks{These authors have equal contribution.}
\author{K. Somesh}
\thanks{These authors have equal contribution.}
\affiliation{School of Physics, Indian Institute of Science Education and Research, Thiruvananthapuram-695551, India}
\author{K. M. Ranjith}
\affiliation{Max Planck Institute for Chemical Physics of Solids, N$\ddot{o}$thnitzer Str. 40, 01187 Dresden, Germany}
\author{M. Baenitz}
\affiliation{Max Planck Institute for Chemical Physics of Solids, N$\ddot{o}$thnitzer Str. 40, 01187 Dresden, Germany}
\author{Y. Skourski}
\affiliation{Dresden High Magnetic Field Laboratory (HLD-EMFL), Helmholtz-Zentrum Dresden-Rossendorf, 01314 Dresden, Germany}
\author{D. T. Adroja}
\affiliation{ISIS facility, Rutherford Appleton Laboratory, Chilton Oxon OX11 0QX, United Kingdom}
\affiliation{Highly Correlated Matter Research Group, Physics Department, University of Johannesburg, Auckland Park 2006, South Africa}
\author{D. Khalyavin}
\affiliation{ISIS facility, Rutherford Appleton Laboratory, Chilton Oxon OX11 0QX, United Kingdom}
\author{A. A. Tsirlin}
\email{altsirlin@gmail.com}
\affiliation{Experimental Physics VI, Center for Electronic Correlations and Magnetism, Institute of Physics, University of Augsburg, 86135 Augsburg, Germany}
\author{R. Nath}
\email{rnath@iisertvm.ac.in}
\affiliation{School of Physics, Indian Institute of Science Education and Research, Thiruvananthapuram-695551, India}
\date{\today}

\begin{abstract}
Magnetism of the spin-$\frac12$ $\alpha$-KVOPO$_4$ is studied by thermodynamic measurements, $^{31}$P nuclear magnetic resonance (NMR), neutron diffraction, and density-functional band-structure calculations. Ferromagnetic Curie-Weiss temperature of $\theta_{\rm CW}\simeq 15.9$\,K and the saturation field of $\mu_0H_s\simeq 11.3$\,T suggest the predominant ferromagnetic coupling augmented by a weaker antiferromagnetic exchange that leads to a short-range order below 5\,K and the long-range antiferromagnetic order below $T_{\rm N}\simeq 2.7$~K in zero field. Magnetic structure with the propagation vector $\mathbf k=(0,\frac12,0)$ and the ordered magnetic moment of 0.58\,$\mu_B$ at 1.5\,K exposes a non-trivial spin lattice where strong ferromagnetic dimers are coupled antiferromagnetically. The reduction in the ordered magnetic moment with respect to the classical value (1~$\mu_{\rm B}$) indicates sizable quantum fluctuations in this setting, despite the predominance of ferromagnetic exchange. We interpret this tendency toward ferromagnetism as arising from the effective orbital order in the folded chains of the VO$_6$ octahedra.
\end{abstract}
\maketitle

\section{Introduction}
Anderson's superexchange theory~\cite{anderson1962} requires that interactions in $3d$ magnetic insulators are predominantly antiferromagnetic (AFM) in nature. Ferromagnetic (FM) interactions are only possible when magnetic ions are close apart, but in this case equally strong AFM interactions will typically occur between more distant atoms, unless special conditions such as charge order~\cite{hasegawa2009,mahadevan2010} or orbital order~\cite{khomskii1973,khomskii2021} are met. In contrast to dozens of well documented material candidates for quantum AFM \mbox{spin-$\frac12$} chains, only a handful of FM spin-chain compounds have been reported. This disparity is in fact not undesirable, because spin-$\frac12$ antiferromagnets show intriguing behavior caused by underlying quantum fluctuations~\cite{quantmag,sachdev2008}, which do not occur in the ferromagnetic case. 

Here, we introduce a material that remarkably departs from this general paradigm. Featuring V$^{4+}$ as the magnetic ion, $\alpha$-KVOPO$_4$ is dominated by short-range FM interactions. It further reveals an antiferromagnetically ordered ground state caused by residual AFM long-range interactions, wherein the reduction in the respective ordered magnetic moment signals quantum fluctuations that appear despite the strong proclivity for ferromagnetism. 

$\alpha$-KVOPO$_4$ belongs to the group of vanadyl compounds with the general formulas $A$VO$X$O$_4$ and $A$V$X$O$_4$F, where $X$ = P or As, and $A$ is a monovalent cation. The common crystallographic feature of this family is the presence of structural chains of the VO$_6$ octahedra linked by corner-sharing. The tetrahedral $X$O$_4$ groups connect the chains into a three-dimensional network, yet leaving large channels for the $A^+$ ions that remain sufficiently mobile, especially at elevated temperatures. This feature triggered interest in $A$VO$X$O$_4$ as potential battery materials~\mbox{\cite{mueller2011,whittingham2014,fedotov2020}}, whereas concurrent magnetism studies have also led to very encouraging results.

From the magnetism perspective, the crucial feature of $A$VO$X$O$_4$ is that their structural chains are not the direction of predominant magnetic interactions, even though the intrachain V--V distances are much shorter than the interchain ones~\cite{Tsirlin144412,Weickert104422}. It has been shown that monoclinic NaVO$X$O$_4$~\cite{Arjun014421,Mukharjee144433} and AgVOAsO$_4$~\cite{Tsirlin144412,Ahmed224423}, as well as triclinic $\varepsilon$-LiVOPO$_4$~\cite{Mukharjee224403}, all adopt an intriguing pattern of crossed bond-alternating spin chains, which are perpendicular to the structural chains. These compounds reveal a field-induced quantum phase transition and an unusual double-dome regime of Bose-Einstein condensation (BEC) of magnons in high magnetic fields~\cite{Weickert104422}. 

\begin{figure}
	\includegraphics{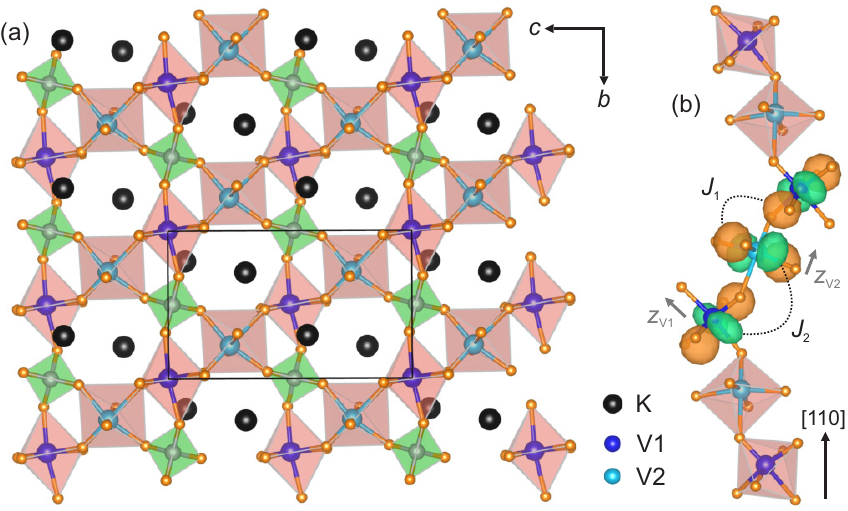}
	\caption{\label{fig:structure}(a) Crystal structure of $\alpha$-KVOPO$_4$ features chains of VO$_6$ octahedra arranged along the $[110]$ direction. (b) Each chain is folded and shows a nearly orthogonal configuration of the magnetic V $d_{xy}$ orbitals because of the different directions of the local $z$-axes (short V--O bonds), which are denoted with the gray arrows. \texttt{VESTA} software~\cite{vesta} was used for crystal structure visualization.
	}
\end{figure}

In the following, we report a comprehensive study of $\alpha$-KVOPO$_4$~\cite{Phillips2158,benhamada1991}, which was so far not on radar of magnetism studies. It is compositionally similar but structurally different (structure: orthorhombic, space group: $P na2_{1}$) from the $A$VO$X$O$_4$ compounds with $A$ = Li, Na, and Ag. The larger K$^+$ ion stabilizes the non-centrosymmetric KTiOPO$_4$-type variety of the crystal structure, where chains of the VO$_6$ octahedra are folded and comprise two nonequivalent vanadium sites, V1 and V2 (Fig.~\ref{fig:structure}). This structural modification has drastic repercussions for the magnetism. Whereas other compounds of the family are AFM in nature, $\alpha$-KVOPO$_4$ is dominated by a FM interaction, yet it shows prominent quantum effects revealed by the reduced ordered magnetic moment. 

\section{Methods}
Polycrystalline sample of $\alpha$-KVOPO$_4$ was synthesized by a solid-state reaction from the stoichiometric mixture of KPO$_3$ and V$_2$O$_4$ (Aldrich, 99.995\%). The KPO$_3$ precursor was obtained by heating KH$_2$PO$_4$ (Aldrich, 99.995\%) for 5 hrs at $300^{\circ}$C in air. The reactants were ground thoroughly, pelletized, and fired at $550^{\circ}$C for 48 hrs in flowing argon atmosphere with intermediate grindings. The phase purity of the sample was confirmed by powder x-ray diffraction (XRD) measurement at room temperature using a PANalytical powder diffractometer (Cu$K_{\alpha}$ radiation, $\lambda_{\rm avg}\simeq 1.5418$~{\AA}).

\begin{figure}
	\includegraphics[width=\linewidth]{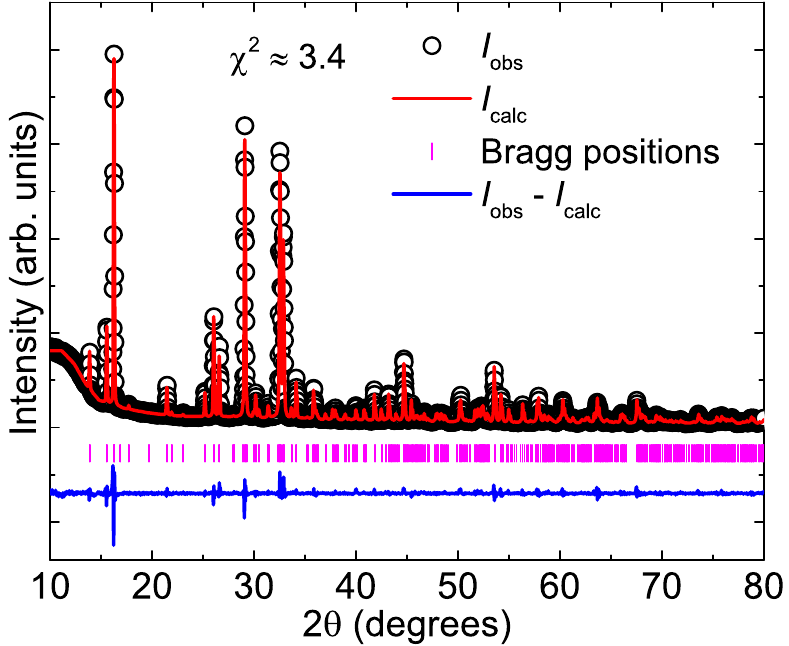}
	\caption{\label{Fig2}Powder x-ray diffraction of $\alpha$-KVOPO$_4$ collected at $T=300$~K. The solid black line denotes the Rietveld refinement fit of the data. The Bragg peak positions are indicated by green vertical bars and bottom blue line indicates the difference between experimental and calculated intensities.}
\end{figure}

Figure~\ref{Fig2} shows the powder XRD pattern at room temperature. Rietveld refinement of the acquired XRD data was performed using the \verb"FULLPROF" software package \cite{Carvajal199355}. The initial structure parameters were taken from Ref.~\cite{benhamada1991}. All the peaks could be refined using Pseudo-Voigt function. The best fit was obtained with a goodness of fit $\chi^2\simeq~3.4$.
The obtained lattice parameters [$a = 12.7612(2)$~{\AA}, $b = 6.3658(2)$~{\AA}, $c = 10.5052(1)$~{\AA}, and V$_{\rm cell}$ = 853.40(4)~{\AA}$^{3}$] are in close agreement with the previous report~\cite{benhamada1991}.

Magnetization ($M$) was measured as a function of temperature (2.1~K~$\leq T \leq$~380~K) and magnetic field ($H$) using the vibrating sample magnetometer (VSM) attachment to the Physical Property Measurement System [PPMS, Quantum Design]. Specific heat ($C_{\rm p}$) as a function of temperature was measured down to $0.35$ K using the thermal relaxation technique in PPMS by varying magnetic field from 0 to 14~T. For $T \leq 2$~K, measurements were performed using an additional $^{3}$He attachment to PPMS. High-field magnetization was measured in pulsed magnetic field at the Dresden high magnetic field laboratory~\cite{Tsirlin132407,Skourski214420}.

The NMR measurements were carried out using pulsed NMR techniques on $^{31}$P (nuclear spin $I=1/2$ and gyromagnetic ratio $\gamma_{N}/2\pi = 17.235$\,MHz/T) nuclei in the temperature range 1.8~K~$\leq T \leq$~250~K. The NMR measurements were done at a radio frequency of $24.76$\,MHz. Spectra were obtained by sweeping the magnetic field at a fixed frequency. The NMR shift $K(T)=[H_{\rm ref}-H(T)]/H(T)$ was determined by measuring the resonance field of the sample [$H(T)$] with respect to the nonmagnetic reference H$_{3}$PO$_{4}$ (resonance field $H_{\rm ref}$). The $^{31}$P spin-lattice relaxation rate $1/T_{1}$ was measured as a function of temperature using the inversion recovery method.

Neutron powder diffraction (NPD) measurements~\cite{Nath_neutron} were carried out at various temperatures down to 1.5\,K using the time-of-flight diffractometer WISH at the ISIS Facility, UK~\cite{Chapon22}. The Rietveld refinements of the NPD data were executed using the \verb"FullProf" software package~\cite{Carvajal199355}.

Density-functional-theory (DFT) band-structure calculations were performed in the \texttt{FPLO} code~\cite{fplo} using Perdew-Burke-Ernzerhof (PBE) flavor of the exchange-correlation potential~\cite{pbe96}. Correlation effects in the V $3d$ shell were taken into account on the mean-field level (DFT+$U$) with the on-site Coulomb repulsion $U_d=3$\,eV, Hund's coupling $J_d=1$\,eV, and double-counting correction in the atomic limit~\cite{tsirlin2011a}. Exchange couplings entering the spin Hamiltonian,
\begin{equation}
 \mathcal H=\sum_{\langle ij\rangle} J_{ij}\mathbf S_i\mathbf S_j
\end{equation}
where $S=\frac12$ and the summation is over bonds $\langle ij\rangle$, were obtained by a mapping procedure from total energies of collinear spin configurations~\cite{xiang2011,tsirlin2014}. Alternatively, we used superexchange theory of Refs.~\cite{mazurenko2006,tsirlin2011b}, as further explained in Sec.~\ref{sec:dft}.

Quantum Monte-Carlo (QMC) simulations were performed using the \texttt{loop}~\cite{loop} and \verb|dirloop_sse|~\cite{dirloop} algorithms of the ALPS package~\cite{alps} on the $L\times L/2$ finite lattices with periodic boundary conditions and $L\leq 36$. The spin lattice used in the simulations is detailed in Sec.~\ref{sec:dft}. 

\section{Results}
\subsection{Magnetization}
\begin{figure}
	\includegraphics{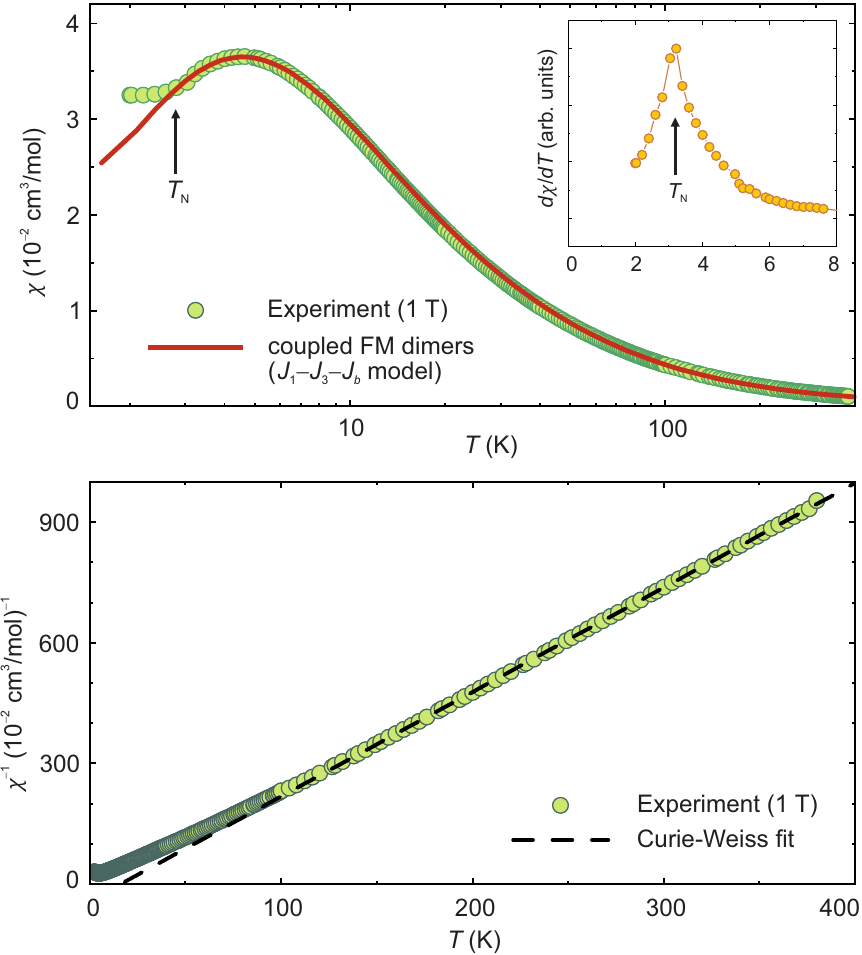}
	\caption{\label{Fig3}Upper panel: magnetic susceptibility $\chi(T)$ measured in the applied field of $\mu_{0}H = 1$~T. The solid line is the fit with the model of coupled FM dimers with $J_1=-150$\,K, $J_3=12$\,K, and $J_b=1.5$\,K (see also Fig.~\ref{fig:magstructure}c and Sec.~\ref{sec:dft}). Lower panel: inverse susceptibility ($1/\chi$) as a function of temperature and the CW fit using Eq.~\eqref{cw}.}
\end{figure}
\begin{figure}
	\includegraphics[width=\linewidth] {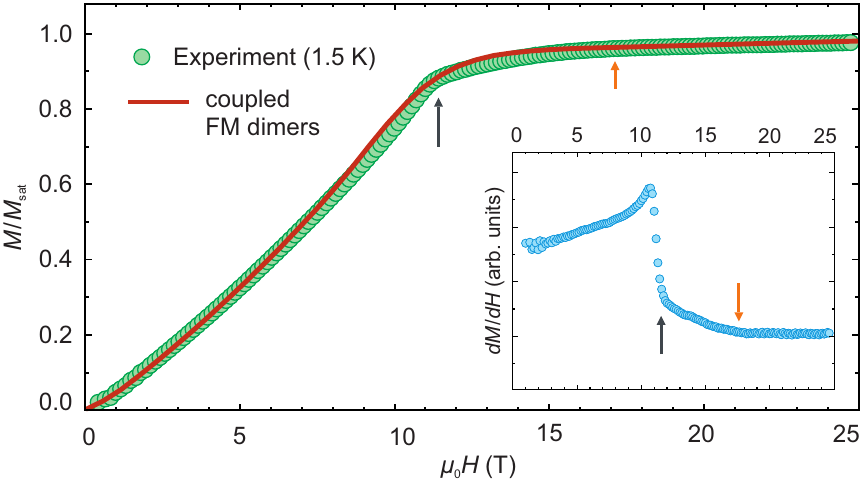}
	\caption{Magnetization vs field measured at $T=1.5$\,K and the fit with the model of coupled FM dimers ($J_1-J_3-J_b$) using the same parameters as in Fig.~\ref{Fig3}. The inset shows $dM/dH$ vs $H$. The colored arrows indicate the field range above $\mu_0H_s\simeq 11.3$\,T where the residual curvature of $M(H)$ is observed (see text for details). }
	\label{Fig4}
\end{figure}
Magnetic susceptibility $\chi(T)~[\equiv M(T)/H]$ measured in an applied field of $\mu_{0}H = 1$~T is displayed in the upper panel of Fig.~\ref{Fig3}. With decreasing temperature, $\chi(T)$ increases in a Curie-Weiss (CW) manner, as expected in the high-temperature regime, and then passes through a broad maximum at $T_\chi^{\rm max} \simeq 5$~K, indicative of an AFM short-range order. A small kink at $T_{\rm N} \simeq 2.7$~K indicates the magnetic ordering transition, which is more pronounced in the ($d\chi/dT$) vs $T$ plot shown in the inset. 
Measurements in the weak applied field of 20\,mT under zero-field-cooled (ZFC) and field-cooled (FC) conditions showed no bifurcation, thus ruling out spin freezing around or below $T_{\rm N}$. In contract to the other $A$VO$X$O$_4$ compounds with the prominent Curie tail below $5-10$\,K~\cite{Tsirlin144412,Mukharjee224403}, the low-temperature susceptibility of $\alpha$-KVOPO$_4$ saturates at a constant value and provides testimony to the high quality of the polycrystalline sample with a negligible amount of paramagnetic impurities and/or defects.

The $\chi(T)$ data in the paramagnetic (PM) region were fitted with the Curie-Weiss law,
\begin{equation}
	\chi(T)=\frac{C}{T-\theta_{\rm CW}}
	\label{cw}
\end{equation}
where $C$ is the Curie constant, and $\theta_{\rm CW}$ is the Curie-Weiss temperature. The fit above 150\,K returns $C \simeq 0.386$~cm$^{3}$K/mol and $\theta_{\rm CW} \simeq 15.9$~K. The $C$ value corresponds to the paramagnetic effective moment of 1.756\,$\mu_{\rm B}$, which is similar to 1.73\,$\mu_B$ expected for spin-$\frac12$. More interestingly, the positive Curie-Weiss temperature signals predominant FM couplings despite the susceptibility maximum above $T_{\rm N}$ and the associated short-range AFM order, which is typical for low-dimensional antiferromagnets.

Magnetization curve measured in pulsed magnetic fields up to 60~T at the base temperature of 1.5~K (Fig.~\ref{Fig4}) approaches saturation around 12\,T. This field defines the energy required to overcome AFM interactions and polarize the spins. It serves as a witness for AFM couplings that should be weaker than the FM ones, though. Interestingly, the magnetization curve is nearly linear. Its curvature is smaller than in the other low-dimensional V$^{4+}$ magnets~\cite{Tsirlin132407,Weickert064403}, including the $A$VO$X$O$_4$ compounds with $A$ = Li, Na, and Ag~\cite{Weickert104422,Arjun014421,Mukharjee144433}. Along with the FM Curie-Weiss temperature, this reduced curvature of $M(H)$ sets $\alpha$-KVOPO$_4$ apart from other V$^{4+}$ materials with the similar composition.

Another peculiarity of $\alpha$-KVOPO$_4$ is its behavior near saturation, where a residual curvature is observed above the kink at $\mu_0H_s\simeq 11.3$\,T before $M(H)$ becomes completely flat around 17\,T (Fig.~\ref{Fig4}, inset). Similar features have been reported in BaCdVO(PO$_4)_2$~\cite{povarov2019,bhartiya2019,skoulatos2019} and assigned to a spin-nematic state expected in a strongly frustrated square-lattice antiferromagnet in the vicinity of saturation~\cite{shannon2006}. However, in the $\alpha$-KVOPO$_4$ case a more trivial explanation, the distribution of saturation fields depending on the field direction, could be an equally plausible reason for the residual curvature above $H_s$. Experiments on single crystals would be interesting in order to pinpoint the exact origin of this feature.


\subsection{Specific Heat}
\begin{figure}
	\includegraphics[width=\linewidth]{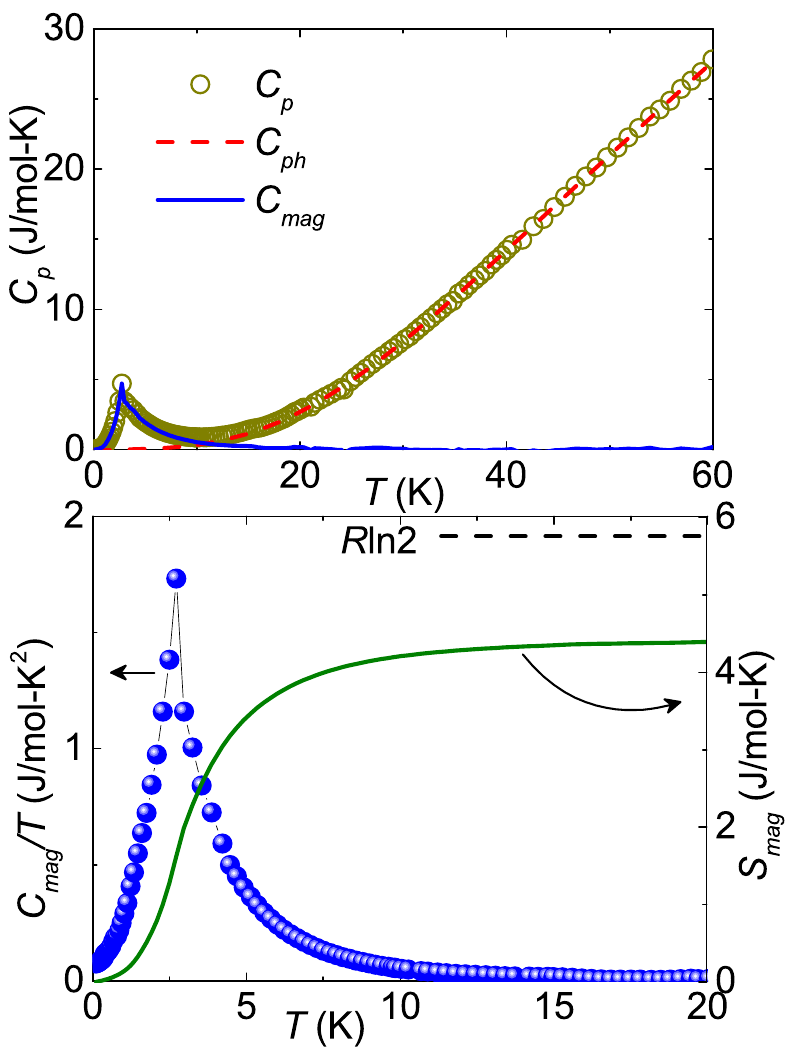}
	\caption{Upper panel: Specific heat ($C_{\rm p}$) vs $T$ for $\alpha$-KVOPO$_4$ in zero applied field. The dashed red line is the phonon contribution to the specific heat ($C_{\rm ph}$) using Debye fit [Eq.~\eqref{Debye}]. The solid blue line indicates the magnetic contribution to the specific heat $C_{\rm mag}$. Lower panel: $C_{\rm mag}$/$T$ and $S_{\rm mag}$ vs $T$ in the left and right $y$-axes, respectively.}
	\label{Fig5}
\end{figure}
\begin{figure}
	\includegraphics[width=\linewidth]{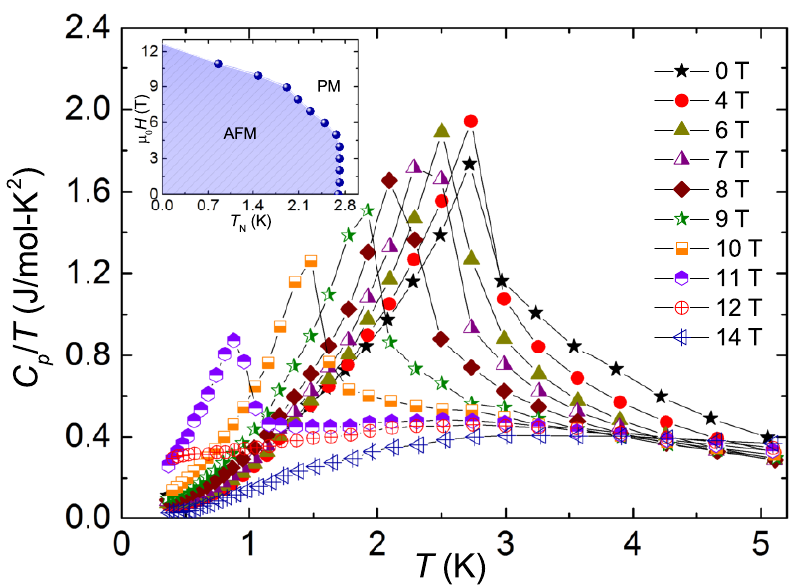}
	\caption{Specific heat divided by temperature ($C_{\rm p}/T$) vs $T$ for $\alpha$-KVOPO$_4$ measured in different fields in the low-$T$ regime. Inset: $H$ vs $T_{\rm N}$ phase diagram.}
	\label{Fig6}
\end{figure}
Specific heat ($C_{\rm p}$) measured in zero magnetic field is shown in the upper panel of Fig.~\ref{Fig5}. In the high temperature region, $C_{\rm p}(T)$ is entirely dominated by phonon excitations, whereas at low temperatures magnetic contribution becomes prominent. The $\lambda$-type anomaly at $T_{\rm N} \simeq 2.7$~K confirms the magnetic transition. 

Magnetic contribution $C_{\rm mag}$ was separated by subtracting the estimated phonon contribution ($C_{\rm ph}$) from the total measured $C_{\rm p}(T)$. To this end, the data above 40\,K were fitted by a linear combination of four Debye functions~\cite{Nath064422,Nath224513}
\begin{equation}
	\label{Debye}
	C_{\rm ph}(T) = 9R\displaystyle\sum\limits_{\rm n=1}^{4} c_{\rm n} \left(\frac{T}{\theta_{\rm Dn}}\right)^3 \int_0^{\frac{\theta_{\rm Dn}}{T}} \frac{x^4e^x}{(e^x-1)^2} dx.
\end{equation}
Here, $R$ is the universal gas constant, the coefficients $c_{\rm n}$ represent the number of distinct atoms in the formula unit, and $\theta_{\rm Dn}$ are the corresponding Debye temperatures. The resulting $C_{\rm mag}$ and the respective magnetic entropy $S_{\rm mag}$ obtained by integrating $C_{\rm mag}(T)/T$ (Fig.~\ref{Fig5}) suggest that about 75\% of the total magnetic entropy $R\ln2 = 5.76$~J/mol~K is released below 10\,K. Additional magnetic entropy should then be released at higher temperatures, but it is difficult to extract because the phonon term becomes predominant. This observation corroborates the scenario of weak AFM couplings that coexist with a much stronger FM exchange. The AFM couplings are responsible for the effects below 10\,K (both short-range and long-range magnetic order revealed by the susceptibility data) and account for the corresponding 75\% of the entropy, whereas the FM coupling is responsible for the remaining 25\% released at higher temperatures. 

We also followed the evolution of $T_{\rm N}$ in the applied field. The transition temperature is nearly unchanged up to 4\,T and decreases gradually in higher fields. Such a behavior is intermediate between classical antiferromagnets, where $T_{\rm N}$ is systematically reduced by the field~\cite{Yogi2736}, and low-dimensional quantum antiferromagnets, where $T_{\rm N}$ initially increases and then becomes suppressed, thus leading to a non-monotonic phase boundary~\cite{sengupta2009,tsirlin2011c}. The absence of such a non-monotonic phase boundary, despite the presence of the susceptibility maximum due to short-range order, also distinguishes $\alpha$-KVOPO$_4$ from a typical low-dimensional antiferromagnet.

\subsection{$^{31}$P NMR}
\subsubsection{NMR Spectra}
\begin{figure}
	\includegraphics [width = \linewidth]{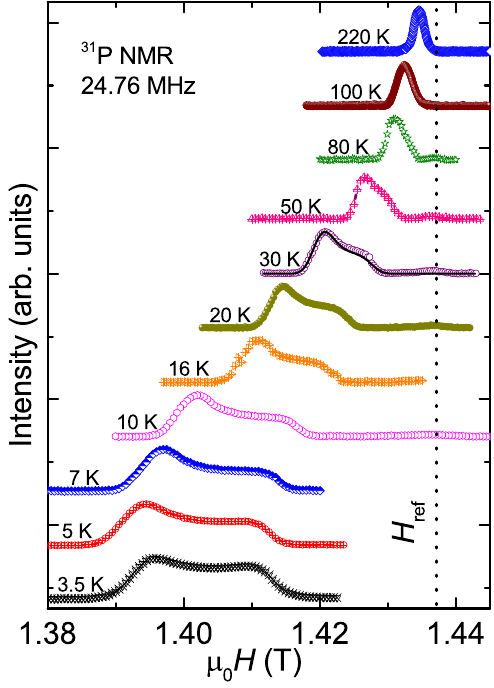}
	\caption{Field-sweep $^{31}$P NMR spectra at different temperatures (for $T > T_{\rm N}$) measured at 24.76\,MHz. The vertical dashed line corresponds to the $^{31}$P resonance frequency of the reference sample H$_{3}$PO$_{4}$. The solid line is the fit to the 30\,K spectrum with the fitting parameters $K_{\rm iso} \simeq 1.03$~\%, $K_{x} \simeq 1.28$~\%, $K_{y} \simeq 1.16$~\%, and $K_{z} \simeq 0.65$~\%.}
	\label{Fig7}
\end{figure}
\begin{figure}
	\includegraphics [width=\linewidth]{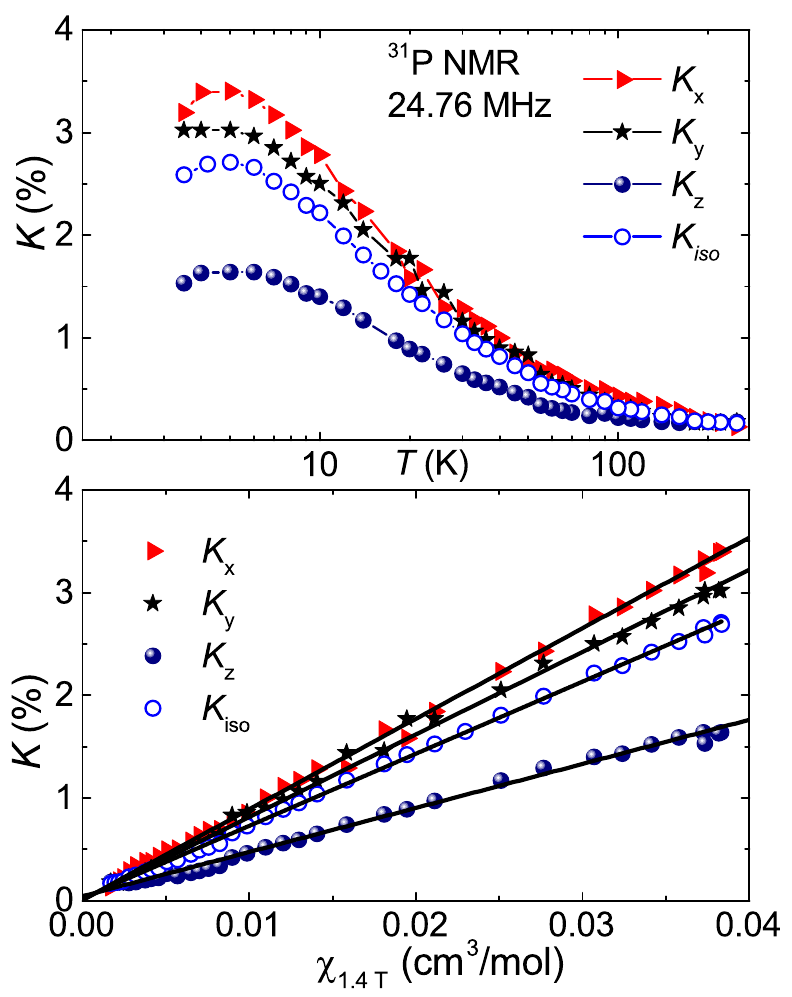}
	\caption{Upper panel: NMR shift components ($K_x$, $K_y$, and $K_z$) and the isotropic shift ($K_{\rm iso}$) as a function of $T$. Lower panel: $K$ vs $\chi$ measured at 1.4\,T is plotted with temperature as an implicit parameter. The solid lines are the linear fits as described in the text.}
	\label{Fig8}
\end{figure}
The $\alpha$-KVOPO$_4$ structure contains two nonequivalent P sites and both of them are coupled to the V$^{4+}$ ions in each chain. Both the P sites reside in an almost symmetric position between the two V$^{4+}$ ions. $^{31}$P being a $I = 1/2$ nucleus, one expects a single and narrow spectral line. 

Figure~\ref{Fig7} presents the $^{31}$P NMR spectra measured at different temperatures. We observed a single spectral line at high temperatures but the line shape is found to be asymmetric, similar to that observed for Zn$_2$VO(PO$_4$)$_2$ and Sr(TiO)Cu$_4$(PO$_4$)$_4$~\cite{Yogi024413,Islam174432}. The single spectral line implies that both the P-sites are almost equivalent.
Further, the asymmetric line shape is likely due to the anisotropy in $\chi(T)$ and/or asymmetric hyperfine coupling between the $^{31}$P nuclei and V$^{4+}$ spins. With decrease in temperature, the line broadens and the asymmetric line shape becomes more pronounced. At very low temperatures, when the system approaches $T_{\rm N}$, the spectra broaden abruptly.

\subsubsection{NMR Shift}
Anisotropic components of the NMR shift ($K$) as a function of $T$ were estimated by fitting each spectrum. The fit of the spectrum for $T = 30$~K is shown in Fig.~\ref{Fig7}. The extracted $K(T)$'s corresponding to three different crystallographic directions ($K_x$, $K_y$, and $K_z$) and the isotropic NMR shift [$K_{\rm iso} = (K_x + K_y + K_z)/3$] are presented in Fig.~\ref{Fig8}. With decrease in temperature, all the components increase in a CW manner and then show a broad maximum around 5~K, similar to $\chi(T)$. Unlike bulk $\chi(T)$, $K(T)$ is insensitive to free spins and defects and allows a more reliable estimate of the intrinsic susceptibility.

The relation between $K(T)$ and $\chi(T)$ is typically written as
\begin{equation}
	K(T)=K_{0}+\frac{A_{\rm hf}}{N_{\rm A}} \chi_{\rm spin}(T),
	\label{shift}
\end{equation}
where $K_{0}$ is the temperature-independent NMR shift, $A_{\rm hf}$ is the hyperfine coupling constant between the $^{31}$P nuclei and V$^{4+}$ electronic spins, and $\chi_{\rm spin}(T)$ is the intrinsic susceptibility.
From the linear relation between $K$ and bulk $\chi$, we determine ($K_{0} \simeq 0.006$~\%, $A^{x}_{\rm hf} \simeq 4927$~Oe/$\mu_{\rm B}$), ($K_{0} \simeq 0.01$~\%, $A^{y}_{\rm hf} \simeq 4487$~Oe/$\mu_{\rm B}$), and ($K_{0} \simeq 0.04$~\%, $A^{z}_{\rm hf} \simeq 2398$~Oe/$\mu_{\rm B}$) for $K_x$, $K_y$, and $K_z$, respectively. The average value of $A_{\rm hf}$ [$= (A^{x}_{\rm hf}+A^{y}_{\rm hf}+A^{z}_{\rm hf})/3$] is calculated to be $\sim 3937$~Oe/$\mu_{\rm B}$ which matches nicely with the $A^{\rm iso}_{\rm hf}$ value ($\sim 3909$~Oe/$\mu_{\rm B}$) obtained directly from the $K_{\rm iso}$ vs $\chi$ analysis. This value of $A^{\rm iso}_{\rm hf}$ is comparable with the values reported for other transition-metal phosphate compounds~\cite{Mukharjee144433,Nath174436,Nath134451}. 

\begin{figure}
	\includegraphics [width=\linewidth]{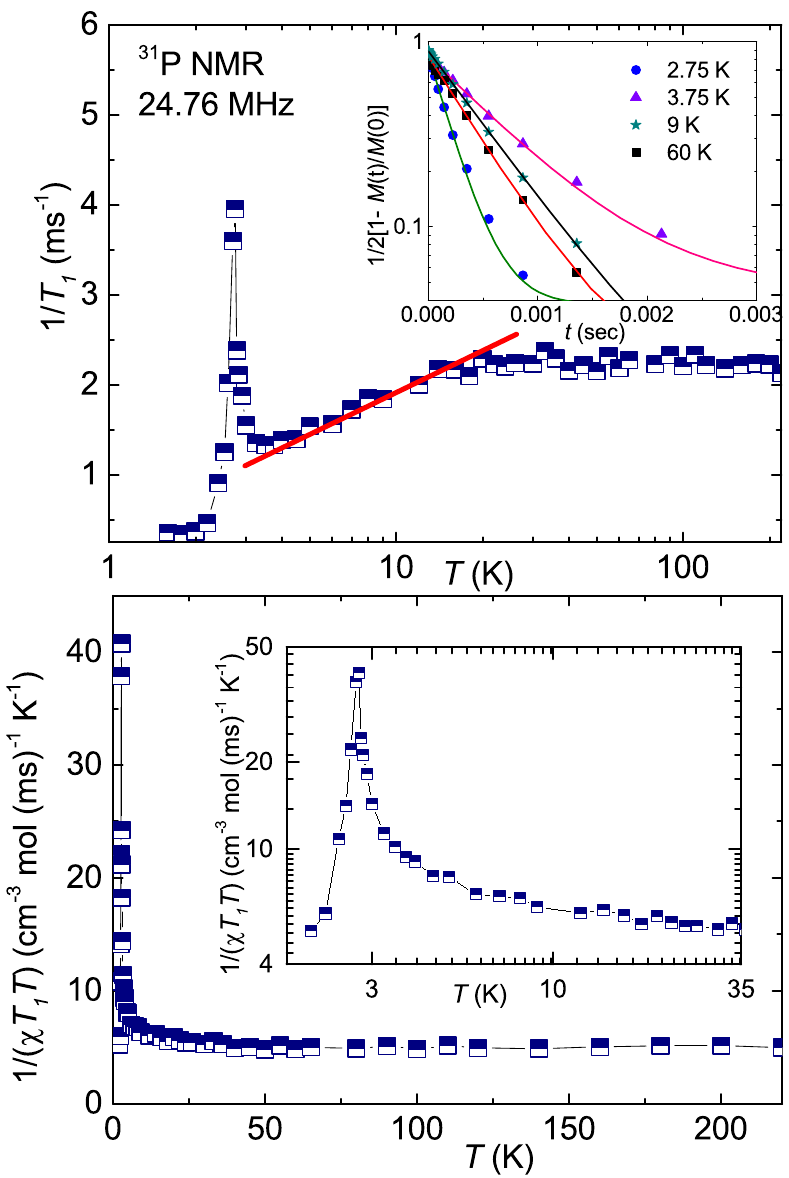}
	\caption{Upper panel: Spin-lattice relaxation rate $1/T_{1}$ vs $T$ measured at 24.96\,MHz. Solid line depicts the linear behaviour for $T \leq 20$~K. Inset: Recovery of the longitudinal magnetization as a function of $t$ for four different temperatures. Solid lines are fits using Eq.~\eqref{exo}. Lower panel: Plot of $1/(\chi T_{1}T)$ vs $T$. Inset: The low-$T$ $1/(\chi T_{1}T)$ data are magnified.}
	\label{Fig9}
\end{figure}
\subsubsection{Spin-lattice relaxation rate $1/T_1$}
The local spin-spin correlations can be understood by measuring the temperature-dependent spin-lattice relaxation rate $1/T_1$, which yields information on the imaginary part of the dynamic susceptibility $\chi(q,\omega)$. The $^{31}$P $1/T_{1}$ was measured by exciting the sample at the field corresponding to the central peak position down to $T = 2$~K. As expected, the recovery of the longitudinal magnetization follows the single-exponential behavior, which is typical for a $I = 1/2$ nucleus. The longitudinal recovery curves at various temperatures were fitted well by the single exponential function,
\begin{equation}
	\frac{1}{2}\left[1-\frac{M(t)}{M({0})}\right]= Ae^{-t/T_{1}} + C.
	\label{exo}
\end{equation}
Here, $M(t)$ is the nuclear magnetization at a time $t$ after the inversion pulse, $M(0)$ is the equilibrium magnetization, and $C$ is a constant. Recovery curves at four different temperatures along with the fit are shown in the inset of the upper panel of Fig.~\ref{Fig9}. The temperature-dependent $1/T_1$ estimated from the above fit is plotted in the upper panel of Fig.~\ref{Fig9}.

As shown in Fig.~\ref{Fig7}, the NMR spectra are asymmetric with two shoulders. In view of this, we measured $1/T_{1}(T)$ at both shoulder positions and found equal values. The data shown in Fig.~\ref{Fig9} correspond to the most intense shoulder position on the left-hand side. With decreasing temperature, $1/T_{1}$ remains constant down to 20~K due to fluctuating paramagnetic moments~\cite{Moriya516}. With further decrease in temperature, $1/T_{1}$ decreases in a linear manner and then exhibits a sharp peak at $T_{\rm N} \simeq 2.72$~K. This sharp peak at $T_{\rm N}$ reflects the slowing down of the fluctuating moments upon approaching $T_{\rm N}$. Below $T_{\rm N}$, $1/T_{1}$ gradually decreases toward zero. 

In the lower panel of Fig.~\ref{Fig9}, $1/(\chi T_{1}T)$ is plotted against $T$. At high temperatures, it is almost temperature-independent and then increases slowly below about 25~K. In the inset of the lower panel of Fig.~\ref{Fig9}, the data near $T_{\rm N}$ are magnified in order to highlight this slow increase. The general expression for $\frac{1}{T_{1}T}$ in terms of the dynamic susceptibility $\chi_{M}(\vec{q},\omega_{0})$ can be written as\cite{Moriya516,Mahajan8890}
\begin{equation}
	\frac{1}{T_{1}T} = \frac{2\gamma_{N}^{2}k_{B}}{N_{\rm A}^{2}}
	\sum\limits_{\vec{q}}\mid A(\vec{q})\mid
	^{2}\frac{\chi^{''}(\vec{q},\omega_{0})}{\omega_{0}},
	\label{t1form}
\end{equation}
where the sum is over the wave vector $\vec{q}$ within the first Brillouin zone, $A(\vec{q})$ is the form factor of the hyperfine interaction, and $\chi^{''}(\vec{q},\omega _{0})$ is the imaginary part of the dynamic susceptibility at the nuclear Larmor frequency $\omega _{0}$. For $q=0$ and $\omega_{0}=0$, the real component of $\chi (\vec{q},\omega_{0})$ represents the uniform static susceptibility ($\chi$). Thus, the temperature-independent $1/(\chi T_{1}T)$ in the high-temperature region ($\geq 30$\,K) indicates the dominant contribution of $\chi$ to $1/T_{1}T$. The slow increase below 25\,K can be attributed to the growth of AFM correlations.
\subsection{Neutron Diffraction}
\begin{figure}
	\includegraphics [width=\linewidth]{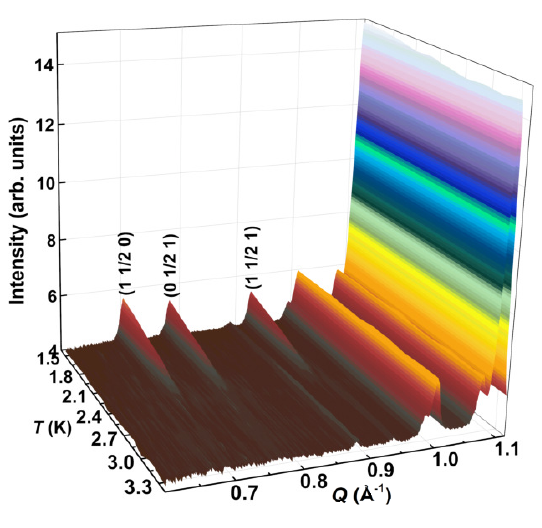}
	\caption{Temperature evolution of the neutron powder diffraction patterns of $\alpha$-KVOPO$_4$ in the low-$Q$ regime and for temperatures below 3.4\,K. An emergence of magnetic reflections with $\mathbf k=(0,\frac12,0)$ is evident below $T_{\rm N}$. }
	\label{Fig10}
\end{figure}
We now probe the magnetic order using neutron diffraction. Figure~\ref{Fig10} shows that three additional magnetic reflections develop below $T_{\rm N}$ at $Q = 0.6973$~{\AA}$^{-1}$, $0.7757$~{\AA}$^{-1}$, and $0.9189$~{\AA}$^{-1}$. They can be indexed with the propagation vector $\mathbf k=(0,\frac12,0)$ suggesting a commensurate magnetic order. 

\begin{figure}
	\includegraphics [width=\linewidth]{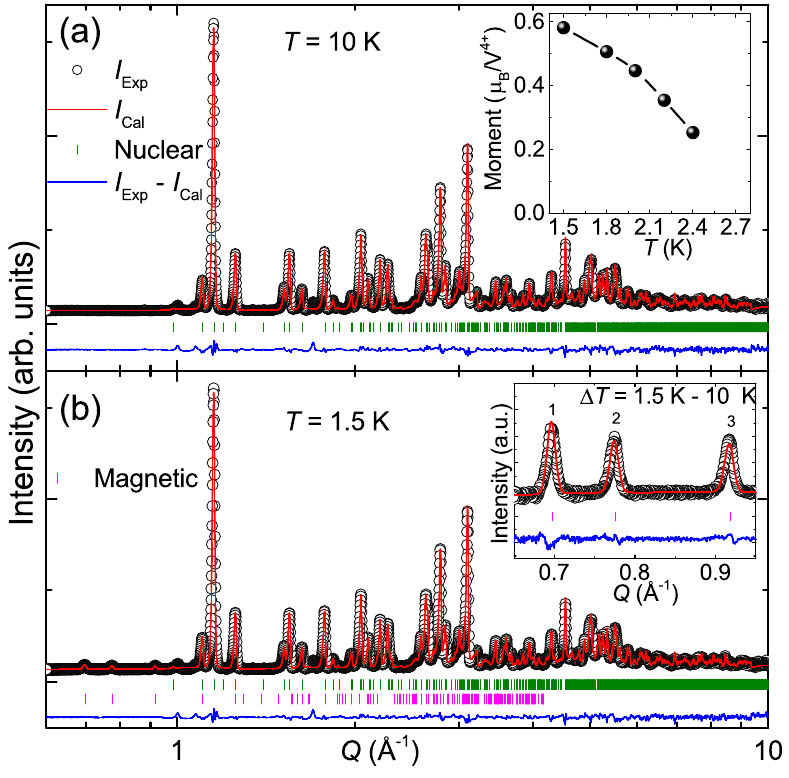}
	\caption{The neutron powder diffraction patterns along with the Rietveld refinement for (a) $T = 10$\,K and (b) $T = 1.5$\,K. Open black circles represent the experimental data, red solid line represents the calculated curve, and difference between them is shown as a blue solid line at the bottom. Vertical marks correspond to the position of all allowed Bragg peaks for the nuclear (top row) and magnetic (bottom row) reflections. The inset in (a) shows the temperature variation of ordered magnetic moment. Inset in (b) presents the Rietveld refinement of the difference data ($\Delta T = 1.5 - 10$~K) using only the magnetic model. }
	\label{Fig11}
\end{figure}

The Rietveld refinement was done using the total diffracted intensities at different temperatures. All the five banks of data have been refined simultaneously to obtain the final parameters. Figure~\ref{Fig11}(a) represents the Rietveld refinement of the nuclear pattern at $T = 10$\,K using the orthorhombic crystal structure with the  space group $P na2_{1}$~\footnote{The vanadium positions were not refined because of the weak coherent scattering by the V nuclei. These positions were thus fixed to the values reported in Ref.~\cite{benhamada1991}. All the atoms sit at the Wyckoff position $4a$. Occupancies of all the atoms were taken to be unity.}.
The obtained lattice parameters are $a = 12.7521(3)$~\AA, $b = 6.3603(2)$~\AA, $c = 10.4985(2)$~\AA, and $ V_{\rm cell} = 851.52(3)~$\AA$^3$.
These values are in close agreement with the refined values from the powder XRD data at room temperature~\cite{benhamada1991}.

\begin{figure*}
\includegraphics{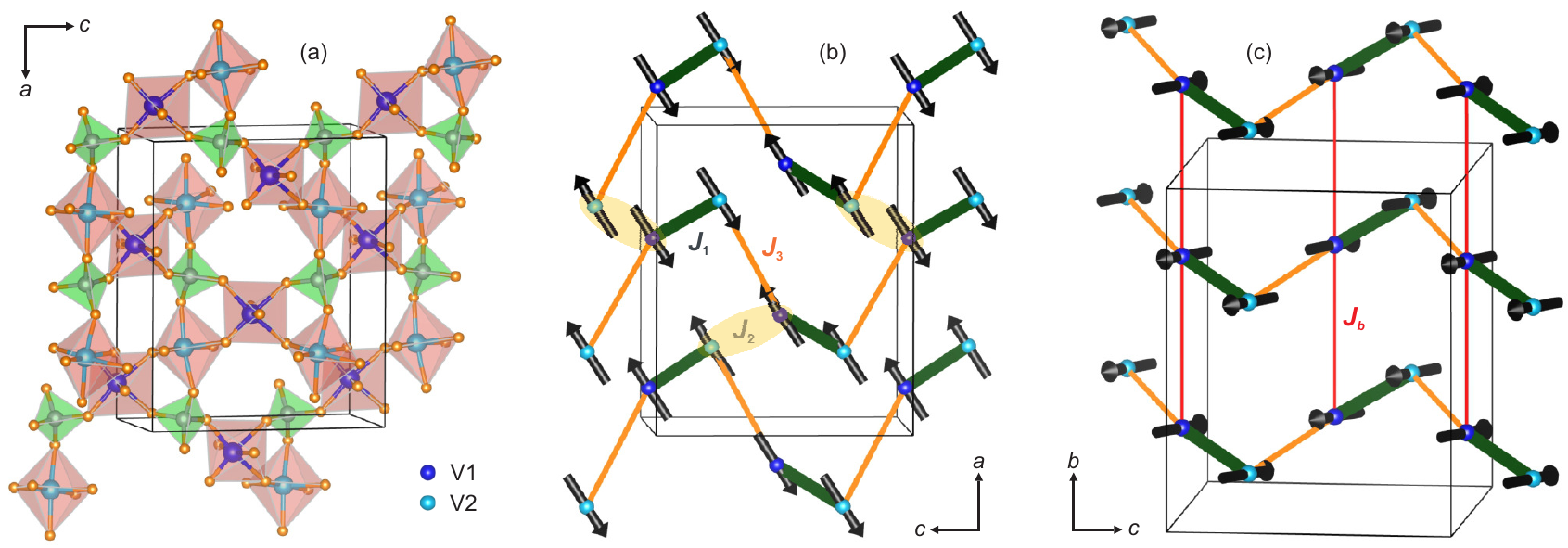}
\caption{\label{fig:magstructure}
(a,b) Crystal and magnetic structures of $\alpha$-KVOPO$_4$ shown in the same projection, with the K atoms  omitted for clarity. The shadings in panel (b) highlight the $J_2$ bonds with the parallel and antiparallel spin arrangement, thus illustrating that $J_2$ does not stabilize the magnetic order. (c) The coupling $J_b$ connects the $J_1-J_3$ chains into layers and causes antiferromagnetic order along $b$ with $\mathbf k=(0,\frac12,0)$.
}
\end{figure*}

Magnetic structure refinement was performed for the diffraction data from the paired 2 and 9 detector banks of the diffractometer because of the best resolution (at an average $2\theta$ value of $58.308\degree$)~\cite{Khalyavin134434}. Figure~\ref{Fig11}(b) shows the combined Rietveld refinement of the nuclear and magnetic reflections at $T = 1.5$~K. A solution was found in the magnetic space group $P\bar 1$. It corresponds to a collinear structure with equal magnetic moments on the V1 and V2 sites. The magnetic moments lie in the $ac$ plane with $\mu_a=0.53(5)$\,$\mu_B$ and $\mu_c=0.23(8)$\,$\mu_B$ at 1.5\,K. The spin arrangement is shown in Fig.~\ref{fig:magstructure} and will be discussed in detail in the next section along with the relevant magnetic couplings.


\subsection{Microscopic magnetic model}
\label{sec:dft}

To analyze magnetic couplings in $\alpha$-KVOPO$_4$, we first consider band dispersions and associated hopping parameters in the band structure obtained on the PBE level without taking strong correlations into account. Band structures of the V$^{4+}$ compounds usually show a well-defined crystal-field splitting imposed by the local environment of vanadium with the short vanadyl bond toward one of the oxygen atoms. This short bond defines the local $z$-direction and renders $d_{xy}$ the lowest-energy, half-filled magnetic orbital. The $d_{xy}$ bands around the Fermi level are typically well separated from the higher-lying bands formed by the four remaining $d$-orbitals~\cite{tsirlin2010,nath2008}.

\begin{figure}
\includegraphics{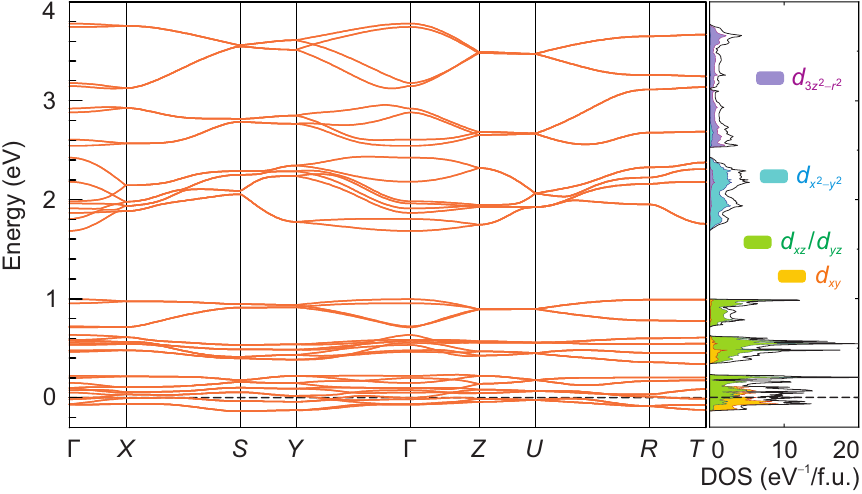}
\caption{\label{fig:bands}
Energy bands and the corresponding density of states (DOS) obtained from the PBE calculation for $\alpha$-KVOPO$_4$. The Fermi level is at zero energy. Note the crystal-field splitting of the V $3d$ states with the significant mixing of $d_{xy}$ and $d_{yz}/d_{xz}$ around the Fermi level.
}
\end{figure}

An inspection of the $\alpha$-KVOPO$_4$ band structure (Fig.~\ref{fig:bands}) suggests a small but important deviation from this conventional scenario. The band complex between $-0.2$ and $0.2$\,eV includes 12 bands per 8 V atoms, so it can't be formed by the $d_{xy}$ states only. Orbital-resolved density of states reveals that these bands include the majority of the $d_{xy}$ contribution, but also a significant portion of the $d_{yz}$ and $d_{xz}$ states. Their admixture is caused by the remarkably large $xy-yz$ and $xy-xz$ hoppings of $0.25-0.30$\,eV for the V--V pairs with the 3.412\,\r A separation ($J_1$ bond).

\begin{table}
\caption{\label{tab:exchange}
Interatomic V--V distances (in~\r A) and the corresponding exchange couplings (in~K) calculated using superexchange theory [$J^{\rm AFM}$ and $J^{\rm FM}$ from Eq.~\eqref{eq:kk}] and mapping analysis ($J^{{\rm DFT}+U}$). The V--V distances are given using the structural parameters from Ref.~\cite{benhamada1991}. The couplings not listed in this table are well below 1\,K.}
\begin{ruledtabular}
\begin{tabular}{ccc@{\hspace{2em}}rrr}
       & $d_{\rm V-V}$ &        & $J^{\rm AFM}$ & $J^{\rm FM}$ & $J^{{\rm DFT}+U}$ \\
 $J_1$ & 3.415         & V1--V2 & 0             & $-154$       & $-149$            \\
 $J_2$ & 3.520         & V1--V2 & 16            & $-9$         & $-3$              \\
 $J_3$ & 5.622         & V1--V2 & 34            & $-4$         & 17                \\
 $J_4$ & 5.731         & V1--V2 & 1             & 0            & $-3$              \\
 $J_5$ & 6.138         & V2--V2 & 4             & $-2$         & 1                 \\
 $J_b$ & 6.360         & V1--V1 & 5             & $-4$         & 1                 \\
 $J_6$ & 6.376         & V1--V1 & 2             & $-3$         & $-2$              \\
\end{tabular}
\end{ruledtabular}
\end{table}

We now use Wannier projections to construct a five-orbital tight-binding model that reproduces all V $3d$ bands, and introduce the extracted hoppings into the superexchange theory~\cite{mazurenko2006,tsirlin2011b} that yields magnetic couplings as 
\begin{equation}
 J=\frac{4t_{xy}^2}{U_{\rm eff}}-\sum_{\alpha}\frac{4\,t_{xy\rightarrow\alpha}^2\,J_{\rm eff}}{(U_{\rm eff}+\Delta_{\alpha})(U_{\rm eff}+\Delta_{\alpha}-J_{\rm eff})}
\label{eq:kk}\end{equation}
where $t_{xy}$ are the hoppings between the half-filled ($d_{xy}$) orbitals, $t_{xy\rightarrow\alpha}$ are the hoppings between the half-filled and empty orbitals, the index $\alpha$ goes through these empty $d$-orbitals, $\Delta_{\alpha}$ are the corresponding crystal-field splittings, $U_{\rm eff}=3$\,eV is the effective Coulomb repulsion, and $J_{\rm eff}=1$\,eV is the effective Hund's coupling. The first and second terms of Eq.~\eqref{eq:kk} stand, respectively, for the AFM ($J^{\rm AFM}$) and FM ($J^{\rm FM}$) contributions to the exchange, as listed in Table~\ref{tab:exchange}.

The large $xy-yz$ and $xy-xz$ hoppings on the $J_1$ bond render the respective magnetic coupling strongly FM. This coupling is augmented by a much weaker AFM $J_3$, whereas all other couplings are below $5-7$\,K in magnitude, either FM or AFM. Our DFT+$U$ mapping analysis (Table~\ref{tab:exchange}) leads to essentially similar results, with the leading FM coupling $J_1$ and the secondary AFM coupling $J_3$. One immediate and rather unexpected consequence of this analysis is that the structural chains with the alternating V--V separations of 3.415\,\r A ($J_1$) and 3.520\,\r A ($J_2$), as shown in Fig.~\ref{fig:structure}b, break into FM dimers formed by the former bond. On the other hand, $J_2$ features weak FM and AFM contributions that nearly compensate each other. Indeed, the FM dimers of $J_1$ can be clearly distinguished in the experimental magnetic structure. In contrast, the spins on the $J_2$ bonds are both parallel and antiparallel [Fig.~\ref{fig:magstructure}(b)], thus indicating that $J_2$ does not contribute to the stabilization of the long-range order.

The two leading couplings, $J_1$ and $J_3$, form alternating spin chains with the strong FM and weak AFM couplings [Fig.~\ref{fig:magstructure}(b)]. Experimental magnetic structure suggests that neither $J_2$ nor $J_4-J_6$ stabilize the order between these chains. The most likely candidate for the interchain coupling is then $J_b$, which runs along the crystallographic $b$ direction and is also responsible for the doubling of the magnetic unit cell [Fig.~\ref{fig:magstructure}(c)]. Our DFT results indeed show a small $J_b$, although between the V1 sites only, whereas the respective interaction between the V2 sites should be below 0.1\,K and thus negligible. The interactions $J_b$ couple the aforementioned alternating spin chains into layers. Long-range magnetic order between these layers may be caused by residual interactions, which are too weak to be resolved in DFT, or by minute anisotropic terms in the spin Hamiltonian.

\begin{figure}
\includegraphics{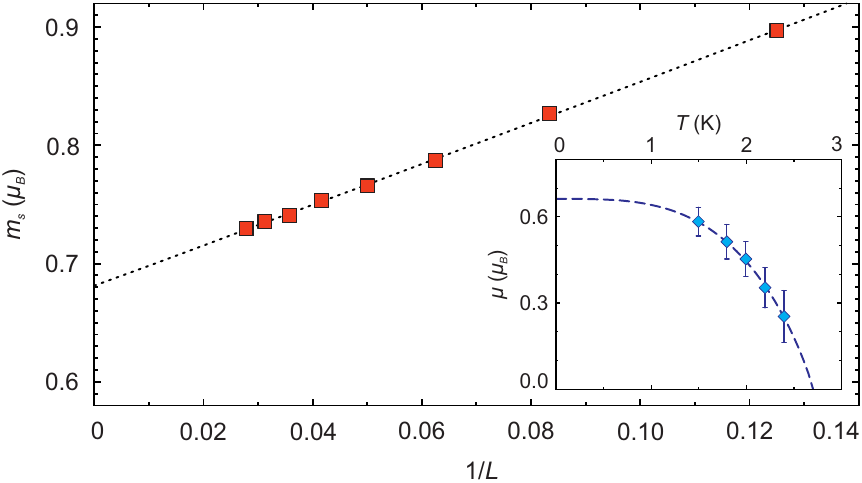}
\caption{\label{fig:moment}
Finite-size scaling of the staggered magnetization ($m_s$) obtained from the static structure factor calculated by QMC at $\beta=1/(k_BT)=16L$. The dashed line is the fit with Eq.~\eqref{eq:ms}. The inset shows the experimental ordered magnetic moment ($\mu$) as a function of temperature and its empirical extrapolation, as described in the text.
}
\end{figure}

While these weak interlayer interactions and/or anisotropy require further dedicated analysis, we argue that the $J_1-J_3-J_b$ model of coupled alternating spin chains reproduces main features of the experimental data. To this end, we simulate thermodynamic properties by QMC and find an excellent agreement with both $\chi(T)$ [Fig.~\ref{Fig3}] and $M(H)$ [Fig.~\ref{Fig4}] using the same set of parameters: $J_1=-150$\,K, $J_3=12$\,K, $J_b=1.5$\,K, and $g=1.98$. Further on, we estimate zero-temperature ordered magnetic moment $\mu_0$ by calculating the spin structure factor $S(\mathbf k)$ for the same set of exchange parameters and using the finite-size scaling for the staggered magnetization~\cite{sandvik1997},
\begin{equation}
 m_s(L)^2= \mu_0^2+\frac{m_1}{L}+\frac{m_2}{L^2}
\label{eq:ms}\end{equation}
where $L$ is the size of the finite lattice, and $m_1$ and $m_2$ are empirical parameters. The resulting $\mu_0\simeq 0.68$\,$\mu_B$ (Fig.~\ref{fig:moment}) reveals the 32\% reduction compared to the classical value of 1\,$\mu_B$. Experimental values of the ordered moment are even lower and affected by thermal fluctuations, because the base temperature of our neutron diffraction measurement is more than half of $T_{\rm N}$. From the empirical scaling, $\mu(T)=\mu_0[1-(T/T_{\rm N})^{\alpha}]^{\beta}$, we estimate $\mu_0^{\rm exp}=0.66$\,$\mu_B$ in an excellent agreement with QMC (Fig.~\ref{fig:moment}, inset).

\section{Discussion and Summary}
%
Weak FM couplings are not uncommon among the V$^{4+}$ compounds. They usually take place between those ions where magnetic $d_{xy}$ orbitals lie in parallel, well-separated planes and lack a suitable superexchange pathway~\cite{Tsirlin144412}. However, more efficient superexchange pathways are always found for other pairs of the V$^{4+}$ ions and give rise to AFM couplings of a similar or even larger strength. $\alpha$-KVOPO$_4$ stands as an exception in this row, because its FM coupling $J_1\simeq -150$\,K is by far the strongest among the known V$^{4+}$ compounds and, therefore, predominant. It arises between the $d_{xy}$ orbitals lying in non-parallel, nearly orthogonal planes and can be traced back to the effective orbital order between the two distinct vanadium sites in the crystal structure, V1 and V2 [Fig.~\ref{fig:structure}(b)]. One interesting question in this respect is why the strong FM coupling is observed for $J_1$ and not for $J_2$, despite the similar V--V distances. This difference can be traced back to the dihedral angles $\psi$ between the planes of the $d_{xy}$ orbitals ($\psi=30.4^{\circ}$ for $J_1$ and $35.5^{\circ}$ for $J_2$) and to the additional superexchange pathways that arise from the PO$_4$ bridges. 

As a system dominated by FM couplings, $\alpha$-KVOPO$_4$ is not expected to show any significant quantum effects. Nevertheless, several experimental observations -- the susceptibility maximum preceding $T_N$, and the reduction in the ordered magnetic moment -- challenge these expectations. Our model of weakly coupled FM dimers reproduces significant features of the experimental data and suggests weak AFM couplings $J_3$ and $J_b$ as the origin of the quantum effects in $\alpha$-KVOPO$_4$. The separation of energy scales into strong FM $J_1=-150$\,K and weak AFM $J_3=12$\,K further implies that at low temperatures an effective description in terms of $S=1$ moments located on the $J_1$ dimers may be appropriate. This description is supported by the magnetic entropy of 4.3\,J/mol\,K released below 10\,K and corresponding to $\frac12 R\ln 3=4.56$\,J\,mol\,K for 0.5 $S=1$ dimers per formula unit. The effective $S=1$ description entails $S=1$ Haldane chains formed by $J_3$ and coupled into layers by $J_b$. Such coupled $S=1$ chains show a quantum phase transition between the magnetically ordered and gapped Haldane phases~\cite{kim2000,wierschem2014}. The proximity to this transition may give a further clue to the reduced ordered moment and quantum effects in $\alpha$-KVOPO$_4$.

In summary, $\alpha$-KVOPO$_4$ is an unusual spin-$\frac12$ magnet featuring strong FM and weak AFM couplings. While the former cause the formation of ferromagnetic spin dimers, the latter connect these dimers into layers and trigger quantum effects. Quantum fluctuations manifest themselves by the short-range order that appears below 5\,K and precedes the long-range order formed at $T_{\rm N}=2.7$\,K. Below $T_{\rm N}$, the commensurate and collinear ground state features a strongly reduced ordered magnetic moment of 0.58\,$\mu_B$ at 1.5\,K. The unusually strong FM coupling $J_1$ is caused by the effective orbital order on the two crystallographically nonequivalent vanadium sites. Overall, from the magnetism perspective $\alpha$-KVOPO$_4$ is entirely different from its siblings, such as NaVOPO$_4$ with its field-induced quantum critical point~\cite{Mukharjee144433} and AgVOAsO$_4$ with the double-dome regime of magnon BEC~\cite{Weickert104422}. The remarkable structural diversity of the V$^{4+}$ phosphates with their different connectivities of the vanadium polyhedra~\cite{boudin2000} suggests that further unusual regimes of magnetic couplings may occur in this broad family of compounds.

\acknowledgments
PKM, KS, and RN acknowledge SERB, India for financial support bearing sanction order no. CRG/2019/000960. We also acknowledge the support of the HLD at HZDR, member of European Magnetic Field Laboratory (EMFL). Thanks to ISIS facility for neutron beam time on the WISH beamline (RB2000205).

%
	
\end{document}